# Empirical Insights from AI-Powered Software Engineering in Practice

**T. Mersch, EKS Intec GmbH, Weingarten, 88250, Germany**

**J. Johns, British Telecom, Ipswich (Suffolk), IP5 3RE, United Kingdom**

**S. Klepke, H. Elhashemy, Siemens AG, Garching b. München, 85748, Germany**

**A. Signell, Vaadin, Turku, 20520, Finland**

**R. Gröpler, ifak e.V., Magdeburg, 39106, Germany**

**K. Schmid, University of Hildesheim, Institute of Computer Science, Hildesheim, 31141, Germany**

***Abstract – Generative Artificial Intelligence (GenAI) is becoming indispensable in future software engineering. Its application spans a wide spectrum - from assisted development to autonomous software engineering, requiring organizations to address associated risks and challenges. This paper investigates these issues through three case studies, highlighting practical problems encountered during GenAI integration. Based on these findings, it provides targeted recommendations for managing risks and ensuring reliable outcomes. This paper provides guidance for effective and responsible GenAI adoption in software engineering based on empirical insights.***

The advancement of Artificial Intelligence (AI) has profoundly influenced numerous domains, with software engineering emerging as one of the affected disciplines. Recent progress in GenAI has demonstrated the potential to transform software engineering workflows. GenAI promises accelerated development cycles by automating repetitive tasks, generating boilerplate code, and refactoring large code bases. Organizations anticipate measurable productivity gains and reduced time-to-market. AI-based systems increasingly contribute beyond code completion, extending towards higher levels of abstraction such as requirement interpretation and system-level reasoning.

The uncritical usage of GenAI in software engineering can lead to erroneous outputs, harmful results, and considerable user frustration. GenAI systems introduce unique risks stemming from probabilistic reasoning, hallucinated content, security vulnerabilities in generated code, and the erosion of fundamental engineering competencies.

In this paper, we examine the spectrum of GenAI utilization in software engineering, ranging from assisted development and task-level orchestration to emerging paradigms such as Vibe Coding and autonomous software agents. We analyze the challenges arising at each maturity level, including issues of trust, verification, governance, skill dependency, and system accountability. Based on multiple industrial and academic case studies, we derive insights and propose strategies tailored to the stages of GenAI integration.

This work provides guidance for effective and responsible GenAI adoption in software engineering.

# AI-POWERED SOFTWARE DEVELOPMENT

A promise of GenAI is that it will enable people to be more productive. Many studies focused on productivity-oriented metrics like task completion time or throughput [1]. An exclusive focus on this can be misleading, as this does not capture second-order effects like an increase in verification effort, sometimes called a "verification tax" [2] nor negative consequences related to cognitive load, communication, etc. [3].

An understanding of the impact of GenAI in software engineering needs to include developer Satisfaction and well-being, Performance, Activity, Communication and collaboration, and Efficiency/flow as argued in the SPACE framework [1]. Such aspects are captured in studies [4], [5].

A problem in evaluating the impact of GenAI on these factors is that it may depend on the specifics of the assistance given by these systems and the details of the development context. This is where our studies contribute: Together, they cover different contexts using GenAI on different assistance levels.

We introduce a model for AI-assisted development. This categorizes the degree of control delegated to autonomous GenAI tools along a single primary axis: *Delegated Autonomy* and over different activities like planning, code generation, and integration. Figure 1 structures approaches to AI-assisted development using this criterion.

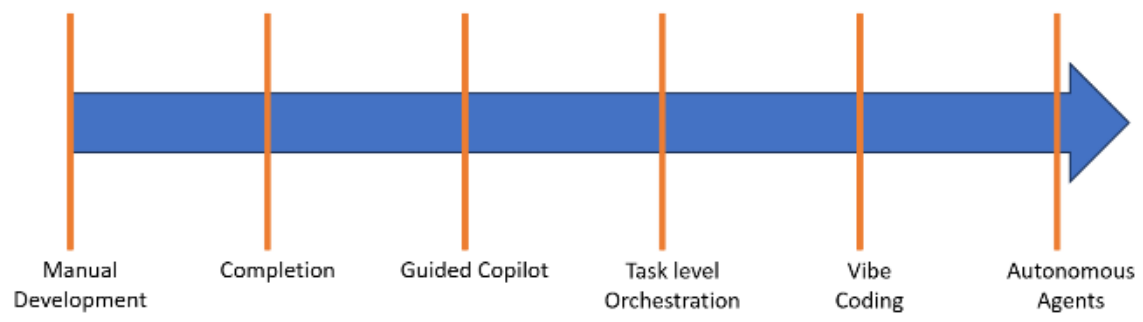

FIGURE 1. Delegated Autonomy Spectrum

Brief definitions of the mentioned paradigms are as follows:

1. **Manual Development. No autonomy.** All development and review activities completed manually by humans.
2. **Assistive Completion. Low autonomy.** GenAI assists with localized tasks (autocomplete, boilerplate generation, refactoring, or documentation) under human direction.
3. **Guided Copilot. Moderate autonomy.** Developers provide natural language instructions at the function or class level; the GenAI drafts code or tests in small diffs, which are reviewed by the developer.
4. **Task Level Orchestration. Higher autonomy.** The developer specifies a goal or ticket; the GenAI proposes a plan, edits multiple files, runs or expands tests, and iterates under human oversight. The human curates and refines rather than specifying each step.
5. **Vibe Coding. High autonomy.** Developers describe high-level intent and constraints; GenAI systems explore implementations, propose integrations, and may push branches or pull requests. Developers primarily orchestrate and adjudicate among alternatives, leaning on CI/testing as guardrails.
6. **Autonomous Agents. Very high autonomy.** Agents initiate tasks, plan, invoke tools (code search, build systems, test runners), and propose merges with minimal human involvement beyond review or escalation.

This spectrum is not intended to prescribe best practice, but to clarify distinctions in the case studies. It mirrors developers' lived experience of fluidity. Engineers commonly move across the spectrum within a session. Key determinants include:

- Codebase maturity and test sufficiency: Well-structured, tested systems allow higher autonomy
- Developer expertise and task fit: Autonomy increases for exploratory or tasks
- Context availability: Rich artifacts (e.g., history, logs, documentation) enable safer delegation
- Risk posture and criticality: Security- or compliance-sensitive areas require tighter control
- Exploration phases favor higher autonomy, while convergence phases reduce it.

Dynamic use of the spectrum is a control mechanism, not a failure and used to balance uncertainty, quality, and velocity. It operates across multiple timescales: Over a system's lifecycle (higher autonomy in prototyping, lower near production) and within individual tasks.

# CASE STUDIES

The ITEA GENIUS Project [6] investigates GenAI across diverse software engineering domains and programming languages. Despite this heterogeneity, analyses revealed recurring challenges across domains and technologies. The following case studies illustrate these challenges in different contexts.

**CaseStudy - EKS (CS1)**

The Virtual Commissioning case study of EKS was conducted in a research team focused on prototyping, where time-to-insight outweighs production readiness, robustness, and maintainability. This motivated the adoption of Vibe Coding. The team served as a pilot environment to generate insights for a larger production-oriented team.

TABLE 1. CS1 Setup.

| | |
|---|---|
| Participant selection: | All team members responsible for prototyping activities were included in the study. The group consisted of developers with heterogeneous experience levels, ranging from junior researchers with limited industrial coding exposure to senior engineers with extensive development backgrounds. |
| Object selection: | To reduce confounding variables, tasks were excluded from participants' routine workload, and an unfamiliar programming language was selected. This prevented dominance of prior domain expertise and enabled clearer observation of GenAI interaction patterns. Developers with varying programming proficiency were deliberately included to analyze differential effects. |
| Type(s) of GenAI used: | The study examined pure Vibe Coding with the help of Claude Sonnet 4.5: Participants avoided manual coding and instead guided GenAI through prompts and iterative refinement to produce functional prototypes. |
| Training provided: | The case study was structured in phases<br><br>Phase 1: No prior formal training was given. Participants relied solely on their intuition and existing experience.<br><br>Phase 2: Informal knowledge exchange sessions were introduced, allowing participants to share prompting strategies and lessons learned.<br><br>Phase 3 (in progress): Structured training modules are being developed and gradually introduced to evaluate the impact of systematic instruction on efficiency and output quality. |
| Study type: | A standardized logbook captured prompts, GenAI responses, iterations, issues, time, quality, and trust, enabling traceable qualitative and quantitative analysis. |

The study addressed the required prior training for AI-assisted development, the optimal assistant type and autonomy level, and developer interaction across varying skill levels.
The team acted as an experimental testbed to derive recommendations for broader organizational adoption.

**Execution**

The empirical evaluation spanned two weeks. During this time, participants engaged in multiple prototyping sessions under controlled but realistic conditions.
Sessions were documented using a prepared logbook. Regular interim reflection meetings were conducted to discuss observations, challenges, and emerging best practices.

**Results**

Vibe Coding accelerates prototyping and high-level design, particularly for UI mockups and initial system and development. GenAI assistants rapidly generated outputs that would otherwise require substantial manual effort, enabling fast iteration.
Limitations emerged in detailed backend development, where GenAI iterations often failed or required manual correction. Token limits disrupted workflows and fostered reliance on AI, indicating cognitive offloading. Users experienced frustration and alienation, as large volumes of generated code and limited language familiarity reduced comprehension. Many participants were unable to explain the produced code, and issues were exacerbated when AI responses overwrote functioning solutions. Users adopted iterative cycles of prompting, testing, and incremental saving.
Vibe Coding was effective for conceptual prototyping but less reliable for fine-grained implementation. Its success depends on stable system access, guidance, and users' ability to critically assess AI-generated outputs.

**Interpretation**

Vibe Coding accelerates prototyping, enabling rapid ideation, architecture design, and scaffolding while lowering cognitive load. It is less reliable for debugging, configuration, and dependency management. Effective use requires critical oversight, validation of outputs, and awareness of automation bias and overreliance.

**CaseStudy – Vaadin (CS2)**

The case study was conducted at Vaadin Ltd, a Finnish software company with 101 employees that develops an open-source Java web framework for building web applications. Multiple development teams contribute to the company's open-source repositories. The study examined how GenAI-assisted development scales across an organization when supported by training and company-wide usage guidelines. Evidence comes from production pull requests, prototyping activities, and internal training.

TABLE 2. CS2 Setup.

| | |
|---|---|
| Participant selection: | Participation was company-wide across product developers in the open-source repositories. Ae GenAI Day workshop attracted 35 participants from both R&D and services departments, including developers working on customer projects.  Experience ranged from daily GenAI users to first-time users encountering AI-assisted development. |
| Object selection: | Two categories of work were included: Production pull requests in open-source repositories and AI-assisted prototypes, proofs-of-concept, and draft pull requests across repositories. A notable example is the AI Orchestrator for the Vaadin Design System, which was prototyped with GenAI assistance and iteratively evolved as specifications and requirements were refined. |
| Type(s) of GenAI used: | Claude Code (Anthropic) served as the primary tool, with ChatGPT and Gemini approved. All tools were used through company-managed accounts configured to prevent model training on submitted data and to automatically delete content according to enterprise retention settings. |
| Training provided: | An internal GenAI Day workshop led by the company's CTOcovered prompt specificity, the importance of planning before coding, context management through repository-level claude.md files, and hands-on issue fixing, applied during a guided coding session. |
| Specific guidance: | Company-wideguidelines defined approved tools, data protection requirements, and personal accountability, while repository-level claude.md files provided project-specific instructions to the agent. Recommendedpractices included working from clean branches, iterative planning before implementation, and GenAI to self-review of its own output. |

**Execution**

Early adoption occurred through individual developers. The December 2025 GenAI Day workshop marked the official starting point. Learnings from the workshop have been brought to planning, prototyping and review assistance with later adoption into the production pipeline. AI Orchestrator became the first experiment where the whole pipeline was utilized. By March 2026, Claude Code had become part of the development workflow.

**Results**

A tracked dataset of AI-assisted pull requests recorded 230 PRs across three repositories as of March 2026: 104 in web-components, 91 in flow-components, and 35 in flow. Of these, 219 were merged contributed by 14 individual developers, excluding automated bot activity.  The workflow followed a consistent pattern: Code was generated with Claude Code, reviewed and revised by the author, reviewed by repository owners, and merged through the standard process. Prototyping of the AI Orchestrator iterated through evolving specifications and proceeded to implementation. AI-assisted workflows surfaced two bugs in an external library,with issues and test cases created with GenAI assistance.

Workshop participants charterized the GenAI as a “mix of a super smart person and a super junior developer, possessing vast knowledge but lacking experience or judgement”, and observed that it “consistently acts like it is the first day at the new workplace, lacking context memory unless explicitly managed”. Vague prompts led to “clown coding” (inefficient cycles of failed attempts), and participants’ recommended approach was to "plan first and code later, often involving iterative planning with the GenAI in planning mode".

**Interpretation**

GenAI adoption scaled because the company provided training and shared guidelines alongside tool access, rather than relying on individual developers to drive uptake: production work, prototyping, training, and policy all progressed. Because GenAI-generated code passed through the same pull request review process human-written code, reviewers remained the quality gate, and the move from experimental to daily use depended on this fit with existing practice rather than on autonomous code generation. The workshop quotes point to a recurring skill gap: effective use depends on prompt specificity and context management, so training plays a central role alongside tool selection. The two-layer guidance model, combining company policy with repository-level instruction files, gives the agent both organisational constraints and project-specific context, making adoption reviewable within the existing development process.

**CaseStudy – Siemens (CS3)**

Since 2024, Siemens has been adopting GenAI in software engineering and collaborating with frontier vendors. For a five-month field study, Siemens choosed to evaluate one of these frontier coding agents - Anthropic Claude Code - focusing on its impact on modernizing large, industrial-grade codebases and for regular development tasks.

TABLE 3. CS3 Setup.

| | |
|---|---|
| Participant selection: | In total 500 Software experts across Siemens volunteered to participate in the field study. The experience level of the participants ranges from Junior (<3 years), advanced (4-7 years) to Senior (10+) years - with 70% of participants being Senior. Most participants have experience in AI-based coding, with some having used Anthropic Claude Code in a preceding pilot study. |
| Object selection: | The study took place from November 2025 to March 2026. Participants used Anthropic Claude Code for their productive code evolution. The tool was used in approx. 20 different industrial software projects, representing a wide range of technologies (web, cloud, desktop). |
| Type(s) of GenAI used: | For the purpose of the study Anthropic Claude Code was chosen to evaluate its impact on evolving and modernizing large industrial-grade codebases. Note, the tool selection is not representative for entire Siemens. |
| Training provided: | Most participants acquired their knowledge about the coding agent through: 1) public sources, e.g. the vendor's website; 2) tool experiment-tation; 3) a channel to share findings to facilitate peer-to-peer learning. |
| Specific guidance: | Participants were provided with company-wide GenAI usage guidelines. |
| Study type: | A five-month, survey-based longitudinal study. The questionnaire (aligned with SPACE framework [1]) comprised 24 questions covering: 1) Tool adoption and use cases, 2) Perceived engineering efficiency, 3) Perceived code quality, 4) Tool suitability for large codebases, 5) Satisfaction, challenges, and trust in generated artifacts, 6) Participants' experience level and average weekly coding time for contextualization. |

**Execution**

Over a period of five months, six anonymous surveys - as described above - were sent out to the 500 participants on a three-weekly basis, yielding 405 total survey responses.

**Results**

An average 76.8% of respondents stated that they frequently or always used the tool during the evaluation period. On average, 84.7% of participants reported being very familiar or moderately familiar with the tool. The top 5 reported use cases were: Code generation (90%), Refactoring (72%), Generation of documentation (67%), Debugging and Bug fixing (63%), Testcases and test data generation (57%).

89.4% of developers reported that the tool contributes significantly or moderately to perceived engineering efficiency. 72.1% reported that the tool contributes significantly or moderately to achieving a state of flow. Data suggests that the self-reported productivity gains are consistent across Junior, Advanced, and Senior Engineers. The same holds true for productivity gains in routine tasks vs. complex tasks. Reported productivity gains were consistent regardless of weekly coding time (<16, 17-32, and 32+ hours).

In the modernization of large code bases, the coding agent excels at understanding large code bases, with 73.3% of feedback being positive. Its effectiveness drops for suggesting (62.2%) and implementing (58.8%) refactoring changes. 73.4% of feedback reported slight or significant improvements in code quality. 50.9% of feedback reported high trust or very high trust in the artifacts generated by Anthropic Claude Code.

Overall, 91.6% reported being satisfied or very satisfied with the tool. Across the evaluation period, 25% reported having challenges with the tool in the following areas:

- Operational friction, e.g., daily manual refresh of security tokens, LLM rate limits, unclear token consumption costs
- Inconsistent performance, e.g., hallucination on complex tasks, especially in large codebases
- Quality and risk-related issues, e.g., overly complex or buggy generated code, requiring significant oversight

Results pertain to code-related tasks, which represent 30-40% of total software project time.

**Interpretation**

The high rates of adoption and familiarity suggest that the tool offers significant, recognizable value for code-related tasks. While the tool demonstrates strong performance in code analysis and understanding, its effectiveness diminishes when tasked with suggesting and implementing changes in large code repositories.

The apparent paradox between satisfaction, perceived productivity, and trust in generated artifacts suggests that while being (very) satisfied with the tool, participants do not blindly trust generated output.

The provided open-ended feedback suggests challenges with the tool in autonomous workflows, and with context engineering best practices for large codebases. Future research must focus not only on improving coding agent and model performance but also on refining the developer experience and adapting to new engineering methodologies.

# CHALLENGES AND RECOMMENDATIONS

Despite spanning distinct regions of the Delegated Autonomy Spectrum, from Vibe Coding (CS1) through task-level GenAI assistance (CS2) to frontier agentic coding at enterprise scale (CS3), the case studies recurring challenges. This section identifies cross-cutting challenges, examines case-specific issues, and derives empirically grounded recommendations.

## Cross-Cutting Challenges

**C1. Verification Overhead and Trust Calibration.** GenAI assistance does not eliminate human judgment, it redistributes it. All three studies showed that accelerated code generation is offset by an increased inspection and validation burden: Evaluation frameworks that measure only generation speed risk overstate net productivity gains.

**C2. Context Management and Prompt Quality.** The stateless nature of large language models requires that all relevant project knowledge, architecture decisions, conventions, and domain constraints be provided in each interaction. Prompt and context engineering should be treated as a first-class engineering competency, not an informal skill acquired through tool exposure.

**C3. Inconsistent Performance Across Task Complexity.** All three studies observed a consistent performance gradient across task complexity. This gradient confirms that appropriate delegated autonomy is dynamically constrained by task complexity and codebase scale, not a fixed property of a team or toolset.

**C4. GenAI Literacy and Training Deficits.** Structured GenAI literacy was absent or insufficient at all three studies, and its development measurably improved outcomes. As GenAI tools are non-deterministic and lack capability boundaries, perceived training deficits may partly reflect tool limitations rather than user knowledge gaps.

GenAI literacy should be treated as an ongoing competency track, not a one-time onboarding step.

**C5. Operational and Tooling Friction.** Tooling friction constituted a source of productivity loss independent of model capability. Effective AI-assisted development depends on the reliability of the surrounding engineering toolchain, an aspect in capability-focused evaluations.

## Case-Specific Challenges

**CS1 - EKS: Cognitive Offloading and Developer Alienation.** Operating at the high-autonomy end of the spectrum, CS1 participants lost the ability to reason about code they had nominally authored. Working in an unfamiliar language, they could not explain AI-generated outputs, and frustration escalated when the GenAI overwrote functional code. The resulting artifacts risk being unownable, a maintenance liability invisible to short-horizon productivity metrics and a significant concern for any organization considering Vibe Coding beyond prototyping contexts.

**CS2 - Vaadin: Organizational Scaling and Governance Infrastructure.** CS2 illustrates that scaling GenAI assistance from individual to organizational use introduces governance challenges. The transition required coordination mechanisms, company-wide usage policies, data protection controls, and repository-level agent instruction files, before GenAI use became reliable and traceable across teams. Responsible organizational adaptions involves compliance and data governance dimensions that extend beyond model selection or individual developer practice.

**CS3 - Siemens: Agentic Workflow Maturity at Industrial Scale.** CS3 surfaced a maturity gap between frontier agent capabilities and the engineering methodologies available: participants encountered the limits of agentic systems on large industrial codebases but lacked established practices for structuring, monitoring, and auditing agentic workflows. The juxtaposition of high satisfaction scores against lower artifact trust reflects a sophisticated user population exercising deliberate skepticism, a posture that must be cultivated and cannot be assumed. At the scale of 500 participants across 20 projects, the cumulative cost of operational friction, though manageable per session, constituted a organizational burden over the five-month study period.

**Recommendations**

**R1. Treat Delegated Autonomy as a Dynamic Governance Variable.** The Delegated Autonomy Spectrum should function as a governance instrument, not merely a taxonomy. The appropriate autonomy level for a task should be calibrated, and periodically revisited, against codebase maturity, test coverage, task complexity, developer expertise, and system risk. Autonomy policies should be integrated with branch protection rules, required checks, and code ownership governance.

**R2. Invest in Context Engineering as a Core Engineering Practice.** Context engineering, the deliberate curation of information provided to GenAI agents, should be treated as a first-class engineering activity alongside documentation and testing. Organizations should budget engineering time for context infrastructure, including well-documented codebases and curated knowledge bases accessible to agents at task time.

**R3. Establish Plan-First Workflows Aligned to Task Complexity.** Define intent before delegating implementation. Requirements and tests provide tasks and correctness guardrails for GenAI agents, while plan-first workflows reduce unproductive iteration through verifiable success criteria.

**R4. Build GenAI Literacy Programs Before Scaling Adoption.** Effective AI-assisted development requires skills in prompt engineering, context management, and agentic workflows beyond tool access alone. Training should be structured, with GenAI competency treated as an ongoing engineering track rather than general onboarding.

**R5. Account for Verification Overhead in Productivity Measurement.** Productivity evaluations should account for verification overhead introduced by AI-generated code, as throughput-only metrics overstate net gains by excluding downstream inspection and integration effort. The SPACE framework [1] and end-to-end PR datasets provide complementary measures, while findings that code-related tasks represent only 30–40% of project time highlight the need to evaluate GenAI within the broader engineering context.

**R6. Enforce Code and Artifact Minimalism.** GenAI agents tend toward unnecessary complexity; teams should enforce code minimalism to support long-term maintenance, performance, security, and accountability.

**R7. Preserve and Develop Domain Expertise Alongside GenAI Adoption.** GenAI amplifies existing strengths and weaknesses; effective use therefore depends on domain expertise rather than tool capability alone.

# CONCLUSION

The most effective GenAI collaborators are domain experts who leverage GenAI as an assistant or delegation tool, requiring both domain-specific education and training in effective GenAI use. Because GenAI amplifies existing strengths and weaknesses, organizations should enforce software engineering best practices and establish governance strategies grounded in risk and cost-benefit analyses. Leadership should define where GenAI is applied, while organizations select appropriate tools and provide targeted training. Sustained investment in domain expertise remains essential for balancing rapid outputs with robust, maintainable systems.

System behavior should be defined through specifications and tests, which provide tasks and guardrails for coding agents. As specifications do not prevent hallucinations, expert review and comprehensive testing remain necessary.

Finally, GenAI agents tend toward unnecessary complexity and artifacts; enforcing code minimalism is therefore essential for long-term maintenance, performance, security, and accountability.

# ACKNOWLEDGMENTS

This work was carried out within the ITEA 4 project GENIUS, as part of the ITEA programme, the Eureka Cluster on software innovation. This work was funded by the German Federal Ministry of Research, Technology and Space (BMFTR) under grant numbers 01IS24069A, 01IS24069D, 01IS24069G, and 16IS24069H, InnovateUK under grant number 600642, and by Business Finland under grant number 3329/31/2024.

During the development of this work, the authors used GenAI to enhance the readability and language of the text. After using these services, the authors reviewed and revised the content as needed. The responsibility for the content of this publication lies with the authors.

**Tina Mersch** is a Research Officer at EKS InTec GmbH, Weingarten, Germany. Her current research interests include end-to-end engineering of production systems, AI integration in engineering processes, and interoperability of industrial data models. She received the Ph.D. degree in engineering from RWTH Aachen University. She is actively involved in standardization initiatives in the field of Industry 4.0 and interoperability across various committees. Contact her at tina.mersch@eks-intec.de.

**Jack Johns** is an AI researcher at British Telecom. He works within BT's Software & AI Lab, researching impact that AI can have on the software development lifecycle. Specific research topics include AI-assisted architectural design, human-AI interaction and knowledge management for multi-agent systems. He holds a BSc in Digital and Technology Solutions from the University of Exeter. Contact him at jack.johns@bt.com.

**Steffen Klepke** is Head of Research Group Development Efficiency and Open Source Software at Siemens Research and Pre-development in Garching / Munich, Germany. His research encompasses Generative AI for industrial-grade Software, industrial-grade DevOps, and Open & Inner Source Software. With his team, he transfers innovation into Siemens business units. He holds a diploma in Business Informatics. Contact him at steffen.klepke@siemens.com

**Hanya Elhashemy**, an AI Senior Technology Strategist at Siemens AG Research and Pre-development. In her role, she defines the company's strategic roadmap to enable the GenAI transformation of industrial software and drives innovative research projects, leveraging GenAI across the entire software development lifecycle. Contact her at Hanya.elhashemy@siemens.com.

**Artur Signell** is CTO and Product Manager at Vaadin, Turku, Finland. His current research interests include efficient and reliable usage of AI for production code, spec-driven development using AI and developer experience tooling for web frameworks. Signell received the M.Sc. (Tech.) degree in Computer Science from Åbo Akademi University. Contact him at artur@vaadin.com.

**Robin Gröpler** is a Research Assistant at ifak - Institute for Automation and Communication, Magdeburg, Germany, and project leader of the ITEA GENIUS project. His work focuses on generative AI for software engineering, particularly in requirements engineering and test automation. He received the Dipl.-Math. degree from Otto-von-Guericke University Magdeburg. Contact him at robin.groepler@ifak.eu.

**Klaus Schmid** is a Full Professor of Software Engineering at the computer science institute of University of Hildesheim, Germany. He works with his research group on software product lines, requirements engineering and software architectures. An important focus of his work is on the use of generative AI in software engineering. He holds a diploma in Computer Science and received a Ph.D. degree from University of Kaiserslautern. Contact him at schmid@uni-hildesheim.de.